# Revealing Rejuvenated Disorder States towards Crystallization in a Supercooled Metallic Glass-Forming Liquid

Wen-Xiong Song[1], Yong Yang[3], Weihua Wang[4], Pengfei Guan[2*]

[1]State Key Laboratory of Functional Materials for Informatics, Shanghai Institute of Microsystem and Information Technology, Chinese Academy of Sciences, Shanghai 200050, China
[2]Beijing Computational Science Research Center, Beijing 100193, China
[3]Department of Mechanical and Biomedical Engineering, City University of Hong Kong, Kowloon, Hong Kong, China
[4]Institute of Physics, Chinese Academy of Sciences, 100190 Beijing, China

We report a metadynamics simulation study of crystallization in a deep undercooled metallic glass-forming liquid by developing appropriate collective variables. Through a combined analysis of free energy surface (FES) and atomic-level behaviors, a picture of an abnormal-endothermic crystallization process is revealed: rejuvenated disorder states with less local fivefold-symmetry and fast dynamics form firstly by changing the local chemical order around Cu atoms, which then act as the precursor for the nucleation of well-ordered crystallites. This process reflects a complex energy landscape with well-separated glassy and crystal basins, giving rise to the direct evidence of intrinsic frustration against crystallization in deep undercooled metallic glass forming liquids. Moreover, the rejuvenated disorder states with distinct physical behaviors offer great opportunities to tailor the performances of metallic glass by controlling the thermal history of a metallic melt.

Since the first metallic glass ($Au_{81}Si_{19}$) [1] fabricated by splat quenching in 1960, the formation of metallic glass through natural cooling processes has stimulated intense research interest [2]. Despite that thousands of bulk metallic glasses (BMGs) [3] have been discovered by the copper mold casting method [4] over the last decade, which exhibited outstanding mechanical and physical properties, such as high strength [5], excellent corrosion resistance [6] and superior soft magnetism [7] etc., the mystery of glass forming ability (GFA) still remains an open issue. The formation mechanisms of BMGs with a very low critical cooling rate have been the long-standing topic of intense discussion [8]. Recent experimental [9] and theoretical [10] results gave hints that the GFA can be investigated from the perspective of frustrated crystallization in supercooled liquids. The competition between glass transition and crystallization during the solidification of supercooled model liquids implies that glass forming can be considered as the result of the frustration of crystallization [11,12]. Thus, simulating the crystallization process in a metallic glass-forming liquid, in the framework of free energy landscape/surface, is of great importance, which could provide the thermodynamic basis for understanding GFA.

Due to its extensive time scale, crystallization in an undercooled liquid, especial good glass-forming liquids, is extremely difficult to simulate using conventional molecular dynamics (MD) simulations and remains as a rare event because of the mismatch between the physical and simulation time. To circumvent this difficulty, advanced sampling techniques such as "persistent embryo" [13], umbrella sampling [14], Monte Carlo (MC) [15,16], and metadynamics [17] (MTD) can be used. With the help of biased potentials, the MTD technique has been employed to map out the free energy barrier for nucleation of water [18], urea [19], pure-element silicon [20], $MgSiO_3$ post-perovskite [21], and $SiO_2$ [22]. According to the theoretical basis of MTD (see the supplemental material [23]), appropriate collective variables (CVs) are the key to reduce the dimension of free energy surface (FES) or potential energy surface (PES) for the concerned states sampling. Unlike the previous successful cases in pure-element or covalent-bond systems [20-22], the crystallization of multi-component metallic system by MTD cannot be driven by rotation strategy (see the SM [23]). It is due to the significantly different local chemical environment between the liquid and crystal phase of multi-component metallic alloys. Consequently, it is critical to develop new CVs to provide extra radial force for the nucleation from multi-component metallic glass-forming liquids.

In this letter, the crystallization of a typical multi-component metallic $Zr_{50}Cu_{50}$ liquid with good GFA was carried out by MTD simulations using embedded atom method (EAM) potential [24], which is implemented in the DLPOLY package [25]. Here, we designed the radical pushing-pulling ($R_{pp}$) parameter as a new CV, defined as $R_{pp}^\alpha = 1 - 1/c_\beta \times \sum_{i=1}^{N_\beta} f_c(r_i) / [\sum_{i=1}^{N_\alpha} f_c(r) + \sum_{i=1}^{N_\beta} f_c(r_i)]$, where the central α-type atom is surrounded by $N_\alpha$ α-type and $N_\beta$ β-type atoms within a cutoff distance $r_{cut}$, $c_\beta$ is the atomic percent of β-type element, and $f_c$ is a switching function to keep coordinate number (CN) changing continuously, to produce the radial force. According to this definition, $R_{pp}^\alpha$ describes the local chemical order (LCO) of α-type atom. For the investigated $Zr_{50}Cu_{50}$ system, α and β could be Cu or Zr element, $c_\beta = 0.5$ and $r_{cut} = 3.64$ Å. It was found that this CV is only effective on the Cu element for accelerating the crystallization. It implies that the LCO of Cu plays an important role for understanding the crystallization of the supercooled $Zr_{50}Cu_{50}$ liquid and will be discussed in detail later. Another CV is selected as Steinhardt order parameter ($Q_6$) to re-construct the bond orientation order (BOO). The details of the two CVs can be found in the SM [23]. The NPT ensemble was applied with Nose-Hoover thermostats with a time step of 2.0 fs in all simulations. The sample size was set up to 686 atoms with 343 Cu atoms and 343 Zr atoms, which is proven to be enough for the crystallization study [23]. In our MTD simulations, an appropriate bias-height $\omega$ is the crucial parameter to strike a balance between accuracy and efficiency for reconstructing the free energy surface (FES). Here, we choose a variable bias-height $\omega = \omega_0 \exp(-V_{aug}/V_{max})$ by adding small potentials close to the end to obtain an accurate FES, where $\omega_0$ equals to 15 $k_B T$, $V_{aug}$ is the current value of the bias energy, and $V_{max}$ equals to 700 $k_B$T. The effective width of the Gaussians augmentation is $\delta h$=0.014 for $Q_6$ and $\delta h$=0.012 for $R_{pp}$ with an interval $\tau_G$ = 2ps.

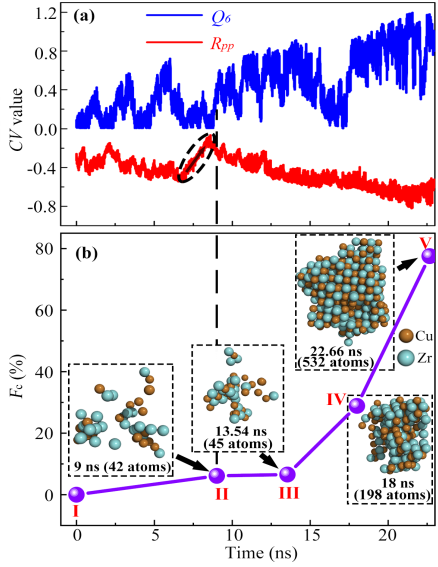

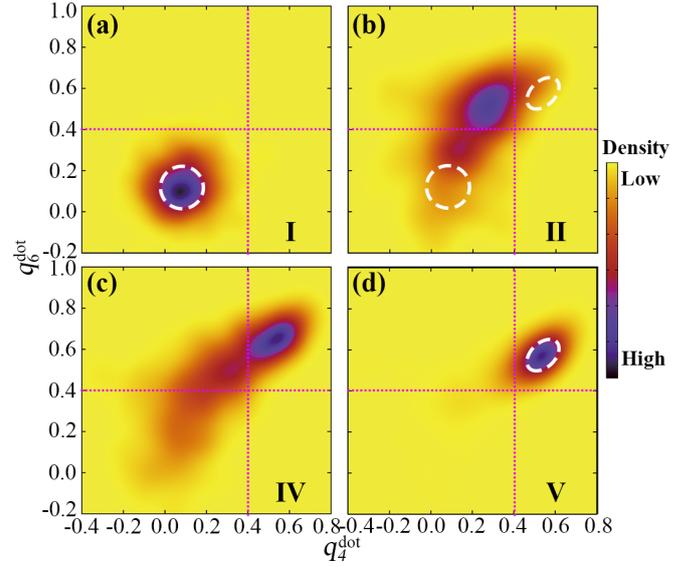

FIG. 1 The crystallization process of $Zr_{50}Cu_{50}$ liquid at 800K (a) The evolution of two CVs, $R_{pp}$ and $Q_6$. (b) The crystallinity fraction as a function of time. The insets are structural snapshots of crystal-like atoms in labeled states. The crystal-like atoms are selected by Voronoi index <0,6,0,8>.

FIG. 2 The distribution of ($q_4^{dot}$, $q_6^{dot}$) for each atom to characterize different crystallization stages of $Zr_{50}Cu_{50}$ MG at 800 K. **(a)~(d)** are the stages of initial nucleation, growth up, and crystallization completing, which are at I, II, IV, and V stages, respectively.

**MTD Simulations.** Figure 1(a) shows the evolution of $R_{pp}$ and $Q_6$ CVs during the crystallization process. The CVs present huge fluctuations during the incubation period, which implies that the initial ZrCu liquid should map out the free energy barrier in a deep free-energy well for nucleation. Remarkably, the $R_{pp}$ presents abnormal evolution in the interval (7ns, 9ns) which implies the obvious change of LCO of Cu atoms. After that, the two CVs perform roughly monotone evolution until the crystallization accomplished. It suggests that the crystallization process can be divided into two distinct ranges by the state corresponding to 9ns. To distinguish the different periods during the process, the fraction of crystallinity ($F_c$) is extracted by Voronoi tessellation analysis [24, 25]. The Voronoi index <i,j,k,l> sequentially represent the number of triangles, tetragon, pentagon, and hexagon in the Voronoi polyhedron. The crystal-like atoms centered at body-cubic-center (bcc) clusters are selected by index <0,6,0,8>. The evolution of $F_c$ is shown in Fig. 1(b). According to classical nucleation theory (CNT) theory, a typical nucleation-growth process can be observed and the size of the critical nucleus is ~45 atoms at 13.54 ns (denoted as III). The transition-state critical nucleus is further confirmed by fifty independent unbiased simulations, where the growth/melt ration is 1.5 to manifest the right point selected. The snapshots of selected bcc-like atoms in four representative states are exhibited in insets. It is clear that the B2-phase is nucleated and grown in $Zr_{50}Cu_{50}$ liquid. Interestingly, the irregular shape of embryos, as observed in the nucleation of water by both MTD [29] and MC [16] simulations, and colloidal system in the experiment [26], may illustrate the anisotropic nucleation originating from atomic-level structural and chemical inhomogeneity. It is noted that, in the state at 9ns (denoted as II), the bcc-like clusters are loosely distributed and totally disappeared as the biased potentials withdraw.

**Local structure evolution.** To quantify the local structure evolution during the crystallization process, an altered bond order parameter "dot-product" $q_l^{dot}$ [27], which utilizes neighboring local environments by averaging the value of dot products of (2$l$+1)-dimension vector of Steinhardt order parameter [34] for all pairs between a central atom and its neighboring atoms, is employed. According to the symmetrical properties of bcc structure, the crystal-like atoms during the crystallization process should be distinguished by $q_4^{dot}$>0.4 associated with $q_6^{dot}$>0.4. Figure 2 summarizes our structure identification results of four states as labeled in Fig. 1. The population of local structures in the ($q_4^{dot}$, $q_6^{dot}$) plane illustrate the structural evolution from initial liquid state I (Fig. 2(a)) to final B2-phase state V (Fig. 2(d)). Unlike the classical crystallization pathway described by the ($q_4^{dot}$, $q_6^{dot}$) plane, population distribution of the intermediate state II is totally different with the initial liquid state I. Usually, the evolution between the population distribution maps of initial state and final state is continuous which is contributed by the interface between the liquid and solid phases during the growth period. The unusual behavior suggests that the state II is a special intermediate state between the initial liquid state and final B2 solid state. It implies that the crystallization pathway is abnormal in $Zr_{50}Cu_{50}$ liquid at 800K.

**Free energy surface (FES).** To investigate the crystallization pathway from the energy landscape perspective, we constructed a CV-constrained FES based on one MTD simulation, as shown in Fig. 3(a). It is clear that the liquid system goes through several basins on the FES with different depths, which encounters some transient liquid states towards the deepest crystalline basin. The FES morphology is similar with but craggier than the schematic one in ref. [28]. Four marked regions I, II, III and V on the FES are corresponding to the labeled states I, II, III and V in Fig. 1(b), respectively. The simplified one-dimensional FES connecting these states is illustrated in Fig. 3(b). It can be found that the states II and III are located at two well-separated sub-basins with non-ignorable energy barriers, but both have a higher free energy than the initial liquid state I, which is consistent with the results in Fig. 2.

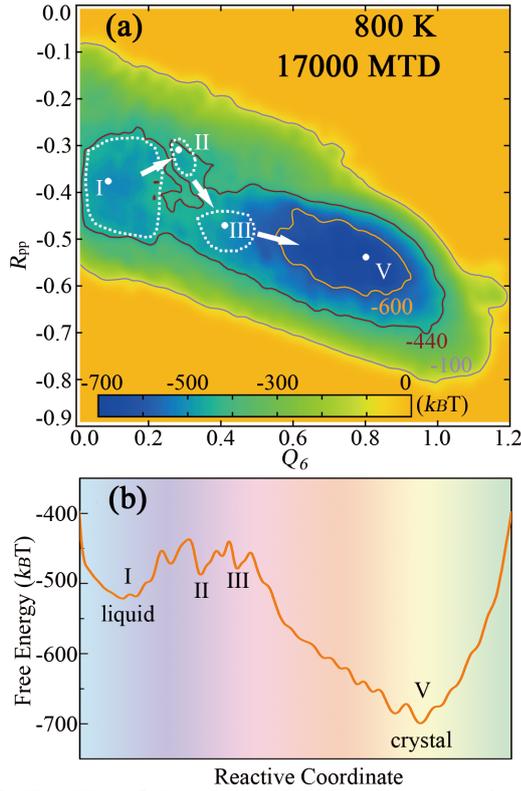

FIG. 3 The FES of $Zr_{50}Cu_{50}$ at 800 K. **(a)**: accumulated FES based on ($R_{pp}$, $Q_6$) CVs. **(b)**: The simplified FES going through the I, II, III, and IV states.

In our MD simulations, the system can stay in the basin II at least 10 ns, which suggests that state II be a rejuvenated intermediate state on the crystallization pathway at 800K. It supposes a complex energy landscape morphology between the well-disorder state and well-ordered state with distinct rejuvenated intermediate states. Since the state III is verified to be the critical nucleation state (Fig 1(b)), thus the state II can act as the precursor has presented in many systems [29-32] to assist the nucleation. Therefore, one may raise an important question: what are the differences between the initial liquid state I and the rejuvenated disorder state II?

**Intermediate rejuvenated disorder states.** As shown in Fig. 1(a) and 3(a), the $R_{pp}$ value dramatically increases as the system evolves from the initial state I to state II, which demonstrates a remarkable change of the LCO of Cu. To characterize the LCO of each element under biased potentials, the calculated CNs and partial PCFs are shown in Fig. S3-4. It was found that the most significant difference between the initial liquid state I and the final crystalline state V is contributed by Cu-Cu pairs, which is the root cause of the effectiveness of $R_{pp}$ on Cu element. We observe the obvious increase of the CN of Cu element associated with the plumping up of the first bottom of the Cu-Cu partial PCF as the system evolves from initial state I to the state II. Since the biased potentials may introduce artificial influences to the system, the actual properties of each state were studied by classical MD simulations.

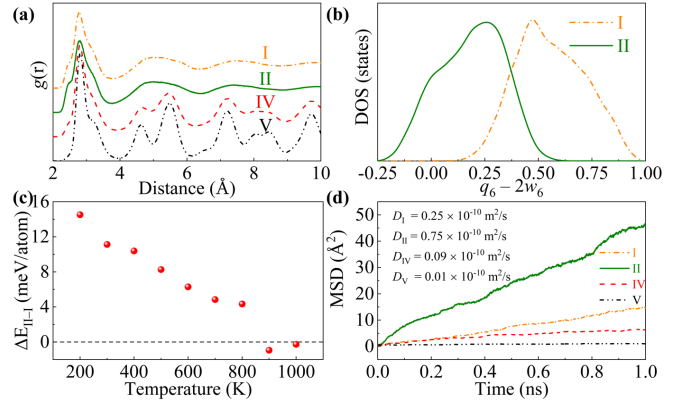

FIG. 4 Different properties of I-liquid and II-liquid states are obtained from unbiased MD simulations. **(a):** Total PCF of four structures at 300 K. **(b):** The distribution of ($q_6$-$2w_6$) value for each atom at 300 K. **(c):** The gradual decrease of total enthalpy difference with the temperature increasing. **(d):** The MSDs of four structures at 800 K for Cu atoms, the inner of which are the corresponding diffusion coefficients.

To characterize the differences between the states of interest without biased potentials, we froze each state to 300K instantaneously and then relaxed 10ns by classical MD simulation for further structure analysis. The calculated total PCFs are shown in Fig. 4(a). The long-range order in the PCFs corresponding to states IV and V indicate the high fraction of crystal-like atoms in states V and IV which is consistent with Fig. 1(b). The PCFs corresponding to states I and II present amorphous behaviors which confirms the structural disorder in state I and II. However, there is no distinguishable difference between the total PCFs of states I and II based on the two-point correlation function.

We further analyzed the local structural information by Voronoi tessellation. The fractions of major Voronoi polyhedron (VPs) are shown in Fig S5. It was found that the factions of both full ICO and ICO-like VPs decrease dramatically in state II by comparing with state I. It suggests that the local fivefold symmetry (LFFS) [33] is weaker in state II. To verify this difference between two liquid states, we utilized the combined order parameter [34] $q_6$–$2w_6$, where $q_6$ is the BOO of each atom and $w_6$ is the reduced invariant one. According to the definition of bond orientation order, the value of $q_6$–$2w_6$ is 1 for full ICO with $q_6 = 0.66$ and $w_6 = –0.17$, but 0 for perfectly homogeneous liquid with $q_6 = 0$ and $w_6 = 0$. Figure 4(b) shows the distribution of $q_6$–$2w_6$ and obvious difference can be observed between states I and II. The blueshift of the peak position illustrates the weaker LFFS feature in the rejuvenated disorder state II than that in initial liquid state I. In previous studies, the LFFS was proposed as the amorphous structural order in supercooled Cu-Zr liquid. This difference between states I and II implies that, to crystallize from a stable amorphous or supercooled liquid, the system should firstly map up the activation barrier by breaking the amorphous order or LFFS. The unstable intermediate II-liquid state encountered probably provides lower diffusion barrier for the inner-shell Cu atom moving out. Recently, the rejuvenated disorder state with less LFFS feature, such as pre-destroying stable polyhedral, has been discussed in deformation-induced crystallization process of MGs [35-38]. For example, the deformation-induced crystallization is simulated at 50 K by cyclic deformation [38], which leads to successive metabasin-to-metabasin transitions by breaking the LFFS. It provides the direct evidence of intrinsic frustration against crystallization in deep undercooled metallic glass forming liquids.

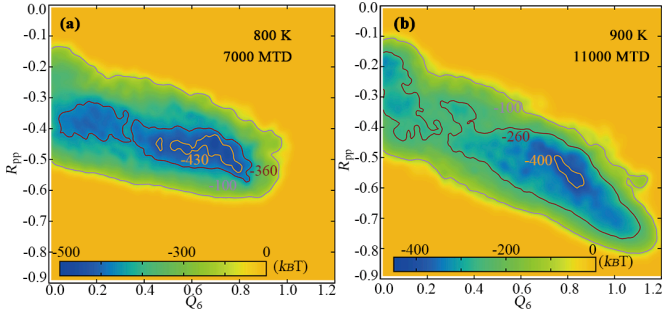

FIG. 5 Controllable crystallization behaviors. **(a):** The FES of II-glass as an initial structure at 800 K. **(b):** The FES of I-glass as an initial structure at 900 K.

The structure difference between glass I and II should be associated with distinct physical properties and the phenomenon of "polyamorphism" recently discussed in the literature [39,40]. As we heat up the glass from 300K to 1000K by MD simulations, the enthalpy difference $\Delta H$ between two systems exhibits a gradual decrease below 800K (Fig. 4(c)). It implies the different thermal capacities of two glasses which is interested for future study. Surprisingly, the $\Delta H$ presents a discontinuous jump to 0 as the temperature increases from 800K to 900K. It gives us the robust evidence that the states I and II located in two distinguishable basins up to 800K. As the temperature higher than 800K, the energy barrier between these two basins is covered by thermal fluctuation. The dynamic behaviors were also investigated by calculating the mean-square displacement (MSD) and diffusion coefficient of investigated states at 800 K. As shown in Fig. 4(d), the state II presents much faster dynamics than other three states. Thus, the intermediate rejuvenated state II, which is thermodynamic metastable at 800K, is a disorder state with lower LFFS and higher diffusivity. As we known, faster diffusion is of great benefit to nucleate because of the existence of inevitable local structure adjustment, such as the change of local chemical order. It indicates that the rejuvenated state acts as the precursor for the nucleation of B2 phase.

**Controllable crystallization pathway.** According to the thermodynamic and kinetic behaviors of state II, we expect to obtain distinct crystallization pathways by controlling the initial state of supercooled liquid. A simpler pathway should be observed if we select state II as the initial state at 800K or perform the MTD simulations at 900K. A simple CV-constrained FES based on the MTD simulation selecting state II as the initial configuration and keeping the same other conditions as the former one is shown in Fig 5(a). The nucleation barrier that must be overcome is 42 $k_BT$ which is about half of the barrier in Fig. 3(a). It can be also verified that the much faster half crystallization time is approximately 10.4 ns. Figure 5(b) shows the accumulated CV-constrained FES during crystallization at 900K. It has a smaller nucleation barrier of 51 $k_BT$ at 900K, where the increased barrier at low temperature originates from the increased radial diffusion barrier by shortening the bond length. On the other hand, if we perform the MTD simulation below 800K, the crystallization cannot be realized in our simulation time scale of 100 ns, where Fig. S6 shows the FES at 600K. All results indicate that the crystallization of MGs can be well controlled by manipulating thermal history systematically, which is powerful to tailor the structure and properties of MG-based alloys.

It should be mentioned that the II state is not only an intermediate state for nucleation, but also an inevitable encountered state for crystal growth. We carry out a 2000-atom MTD simulation and surprisingly find that the crystallization proceeds through many steps that the first step forms a small crystallite of ~500 atoms of no growth up further, and then the crystallite melt and re-crystallize completely, as shown in Fig. S7. It is caused by partial liquid region transferred to the II-like state firstly, which demonstrates that the growth of the crystallite should proceed through the rejuvenated state. It is manifested by the much slower growth rate for ZrCu glass former than the poor glass former NiAl [10]. It also clearly illustrates that the 686-atom system choose here is large enough for the crystallization study in the deep undercooled temperature.

In summary, the simulated crystallization process in a deep undercooled ZrCu MG reflects a complex energy landscape with well-separated glassy and crystal basins, giving rise to the direct evidence for the intrinsic frustration against crystallization in the deep undercooled metallic glass forming liquid. A physical picture of abnormal-endothermic crystallization process can be inferred through the breakage of ICO-like local patterns, which leads to rejuvenated disorder states as the intermediate liquid states with less local fivefold-symmetry and fast dynamics. These intermediate liquid states act as kinetic/thermodynamic barriers against the overall crystal nucleation. From the structural perspective, this is mainly achieved by the directional diffusion of Cu element, but not Zr element, for regulating the change of local chemical order, which also dominates the crystallization of the $Zr_{50}Cu_{50}$ MG. Moreover, the rejuvenated disorder states with distinct physical behaviors on the FES offer a great opportunity to tailor the performance of MGs by controlling the thermal history of the corresponding metallic melts.

We gratefully acknowledge Prof. Limin Liu (Beijing University of Aeronautics and Astronautics) for fruitful discussions. This work was supported by the NSF of China (Grant No. 51571011, U1930402). W.S. acknowledges the support by National Natural Science Foundation of China (61904189). YY is supported by the General Research Fund, Research Grants Council (RGC), through the grants CityU11213118 and CityU11209317. We acknowledge the computational support from the Beijing Computational Science Research Center (CSRC).

Supplemental Material

# Revealing Rejuvenated Disorder States towards Crystallization in a Supercooled Metallic Glass-Forming Liquid

Wen-Xiong Song[1], Yong Yang[3], Weihua Wang[4], Pengfei Guan[2*]

[1]State Key Laboratory of Functional Materials for Informatics, Shanghai Institute of Microsystem and Information Technology, Chinese Academy of Sciences, Shanghai 200050, China

[2]Beijing Computational Science Research Center, Beijing 100193, China

[3]Department of Mechanical and Biomedical Engineering, City University of Hong Kong, Kowloon, Hong Kong, China

[4]Institute of Physics, Chinese Academy of Sciences, 100190 Beijing, China

I.     **Methods;**

   a. Collective Variables;

   b. The $q_l^{dot}$ order parameter;

   c. MD simulations.

II.    **Free energy;**

III.   **Self-diffusion coefficient;**

IV.    **Trajectories travelled on FES every 2 ns in [0, 24] ns region;**

V.     **Accumulated trajectories travelled on FES with the interval of 2 ns in [0, 36] ns region;**

VI.    **Structural evolution of the coordinate number and pair correlation function;**

VII.   **The main polyhedral distributions in I and II states;**

VIII.  **The simulated FES of $Zr_{50}Cu_{50}$ at 600 K;**

IX.    **System size effect of meta-dynamics simulation.**

# I. Methods

## a. Collective Variables (CVs)

Metadynamics (MTD) method aims to characterize the free energy surface (FES) in terms of one or more collective variables (CVs) [1,2]. The identification of appropriate CVs is the key to carry out MTD simulations. Although we do not impose a nucleation pathway during the simulation, the CVs naturally restrict the available pathways to the special one that it can be capable of describing. In order to characterize the local structure order or disorder, Steinhardt order parameter $Q_l$ is used, which is a rotation term. From our previous studies in CuZr system [3,4], the local structure is inhomogeneous. We define a new CV, named as radical pushing-pulling ($R_{pp}$) parameter, to control the local chemical order through the radical movement of atoms.

The general Steinhardt order parameter $Q_l$ of a continuous version is defined as:

$$Q_{lm}^{\alpha\beta} = \left[\frac{4\pi}{2l+1} \sum_{m=-l}^{l} \left|\frac{1}{N_\alpha N_\beta} \sum_{b=1}^{N_\beta} f_c(r) Y_{lm}(\theta_b, \phi_b)\right|^2\right]^{\frac{1}{2}}, \quad (1)$$

where the spherical harmonics describes the local environment of the central $\alpha$-type atom surrounded by the $\beta$-type atoms. The summation in the equation runs over all $N_\beta$ $\beta$-type atoms surrounding the $\alpha$-type atoms within the $r_c$ cutoff ($r_c$ = 3.64 Å herein); $N_\alpha$ presents the total number of $\alpha$-type atoms; $r$ represents the scalar distance between two atoms. The function $f_c(r)$ is a switching function that sets the cutoff range at the required separation in a continuous (and therefore differentiable) manner. It has the form,

$$f_c(r) = \begin{cases} 1, & r \leq r_1 \\ \frac{1}{2}\left\{\cos\left[\frac{\pi(r-r_1)}{r-r_2}\right] + 1\right\}, & r_1 \leq r \leq r_2 \\ 0, & r > r_2 \end{cases}, \quad (2)$$

where the parameters $r_1$ and $r_2$ define a range over which the atom gradually ceases to count. Here, $r_1$ and $r_2$ are set as 2.80 Å and 3.64 Å, respectively.

The local chemical order is continuously controlled by $R_{pp}$ parameter, where the atoms cooperatively move to,

$$R_{pp} = 1 - \frac{\sum_{\beta=1}^{N_\beta} f_c(r)}{\left[\sum_{\alpha=1}^{N_\alpha} f_c(r) + \sum_{\beta=1}^{N_\beta} f_c(r)\right] c_\beta}, \quad (3)$$

where all the parameters in the equation are in line with the previous and $c_\beta$ is the atomic percent of type $\beta$. It illustrates that all $\alpha$-center atoms are surrounded by $\beta$-type atoms as $R_{pp}$ is equal to $\sim(1 - 1/c_\beta)$; and all $\alpha$-center atoms are surrounded by the same $\alpha$ type atoms as $R_{pp}$ is $\sim 1$; and $\alpha$-center atoms are surrounded by some $\alpha$-type atoms and some $\beta$-type atoms as $R_{pp}$ is $\sim 0$.

The embedded atom method (EAM) potential [5] is used in the MTD simulations, which is implemented in the DLPOLY molecular dynamics package [6]. The MTD simulations sample the isothermal-isobaric ensemble (*NPT*) under 1 atm pressure condition with a time step of $\Delta t = 2$ fs. Temperature and pressure are regulated with a Nosé–Hoover thermostat coupled to an isotropic barostat. All results reported use Gaussian augmentations added at intervals of $\tau_G = 2$ ps with a height of 15 $k_B T$. The effective width of the Gaussians is $\delta h=0.014$ in $Q_6$, $\delta h=0.012$ in $R_{pp}$.

## b. The $q_l^{dot}$ order parameter

Steinhardt's order parameter is defined in the equation (1), which is often used to descript the global crystalline order. A local version of $Q_l$ can be defined for each atom in the following vector:

$$\boldsymbol{q}_l(i) = \begin{pmatrix} q_{l,l} \\ q_{l,l-1} \\ \cdots \\ q_{l,-l+1} \\ q_{l,-l} \end{pmatrix} = \left(q_{lm}(i)\right)_{m=-l,l}, (4)$$

$$q_{lm}(i) = \frac{1}{N_i} \sum_{j \in \Omega_i} f_c(r) Y_{lm}(ij), (5)$$

The switching function $f_c(r)$ is introduced to smooth the change of coordinate number.

The norm of $\boldsymbol{q}_l(i)$ is a local $Q_l$ version for an atom:

$$q_l(i) = \sqrt{\frac{4\pi}{2l+1}} \|\boldsymbol{q}_l(i)\|, (6)$$

The order parameter $q_l^{dot}$ is defined based on the bond order correlation $C_{ij}$ between neighboring atoms, first introduced by Frenkel and his coworkers. [7]

$$C_{ij} = \frac{\boldsymbol{q}_l(i) \cdot \boldsymbol{q}_l^*(j)}{\|\boldsymbol{q}_l(i)\| \cdot \|\boldsymbol{q}_l^*(j)\|}, (7)$$

The order parameter $q_l^{dot}$ is the averaged sum of the bond order correlation $C_{ij}$, which is defined as the dot product of $\boldsymbol{q}_l(\boldsymbol{i})$ and the complex conjugate of $\boldsymbol{q}_l(\boldsymbol{j})$, divided by the rotationally invariant norm

of the two vectors:

$$q_l^{dot}(i) = \frac{1}{N_i} \sum_{j \epsilon \Omega_i} f \cdot C_{ij}, (9)$$

**c. MD simulations**

During virtual sample preparation, it was first melted and equilibrated above 2000 K and then cooled to 50 K with a cooling rate of $1 \times 10^{13}$ K/s, during which the cell size is adjusted. Before MTD simulations, the virtual samples are relaxed 1 ns by general molecular dynamics simulations with *NPT* ensemble at a corresponding temperature at 800 K. 686 atoms are used in the virtual sample.

## II. Free energy.

According to the history of adding bias Gaussian potentials, we accumulate these potentials using the following formula:

$$V[s^M(r^N, t)] = \sum_{i=1}^{N_G} \omega \prod_{j=1}^{M} exp\left(\frac{-|s^j(i\tau_G) - s^j(t)|^2}{2\delta_h^2}\right). \quad (10)$$

where $V$ is a time dependent bias potential, which sum over $N_G$ products each of which is a product of $M$ exponents; $\tau_G$ is interval time; $M$ is the number of CVs and $N_G = (t/\tau_G)$; $s^M$ is the value of CVs; $\omega$ and $\delta_h$ are Gaussian weight and width.

## III. Self-diffusion coefficient.

The self-diffusion coefficient $D$ in liquids is obtained by fitting mean-square displacement (MSD) formula after structure relaxed 1 ns under the *NPT* ensemble. MSD is defined as $<r^2(t)> = <|\mathbf{r_i}(t) - \mathbf{r_i}(0)|^2>$. Here $\mathbf{r_i}(t)$ is the position of $i^{th}$ atom at time $t$ and $<>$ denotes ensemble average. The diffusion coefficient $D$ can be calculated by $D = \frac{1}{6} \lim_{t \to \infty} \frac{<r^2(t)>}{t}$.

## IV. Trajectories travelled on FES every 2 ns in [0, 24] ns region.

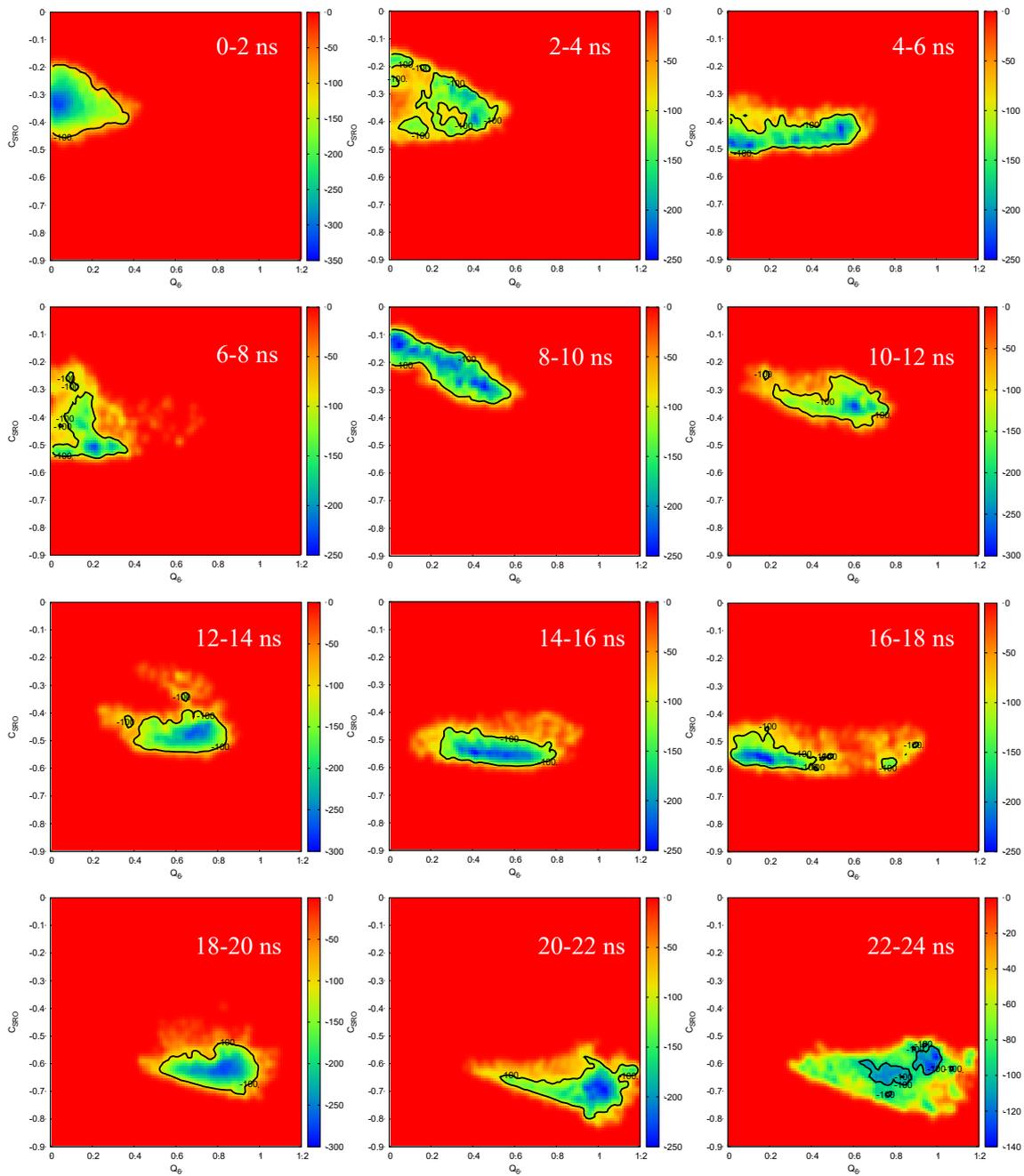

Fig. S1 The trajectories travelled on FES every 2 ns in [0, 24] ns region.

**V. Accumulated trajectories travelled on FES with the interval of 2 ns in [0, 36] ns region.**

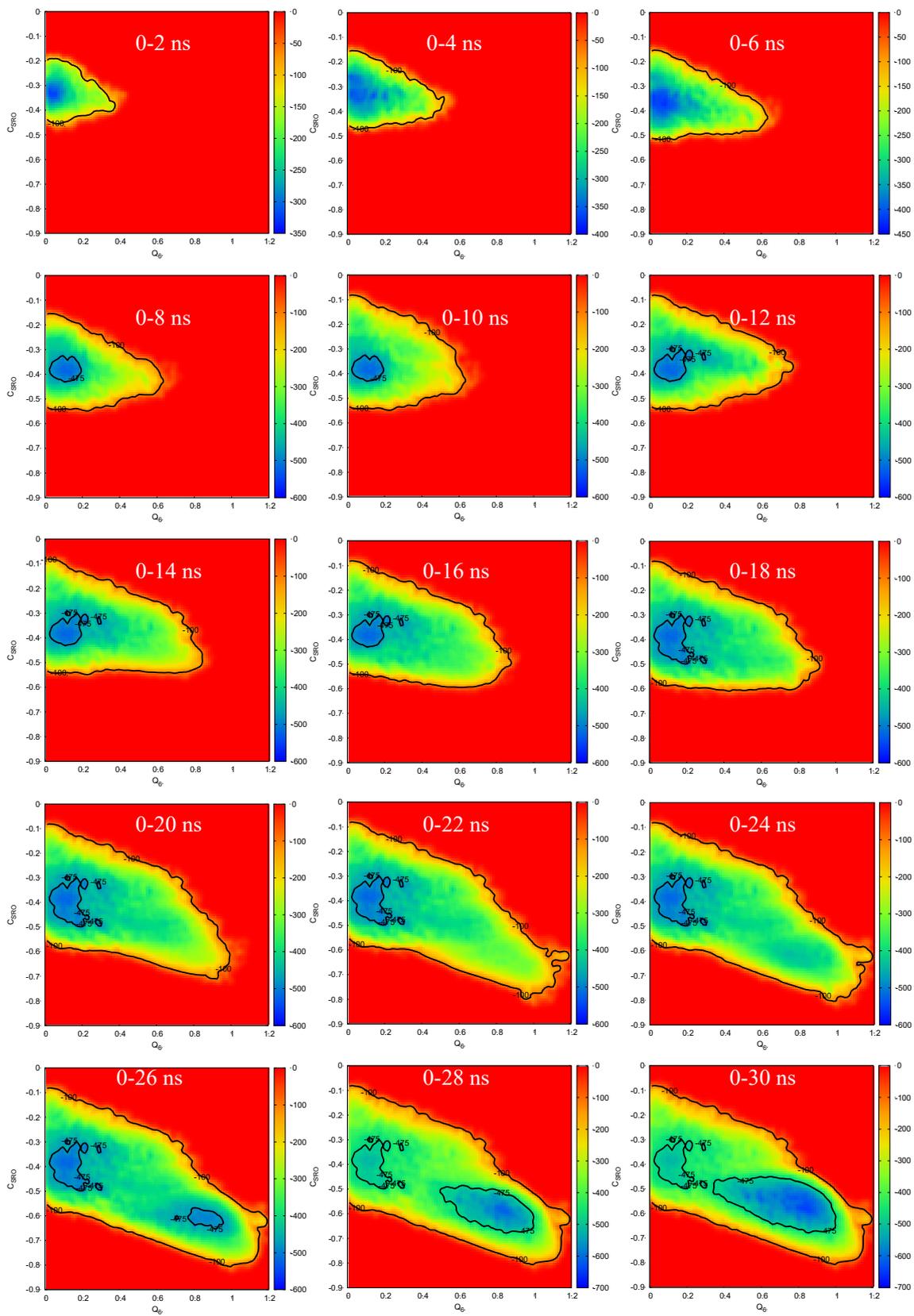

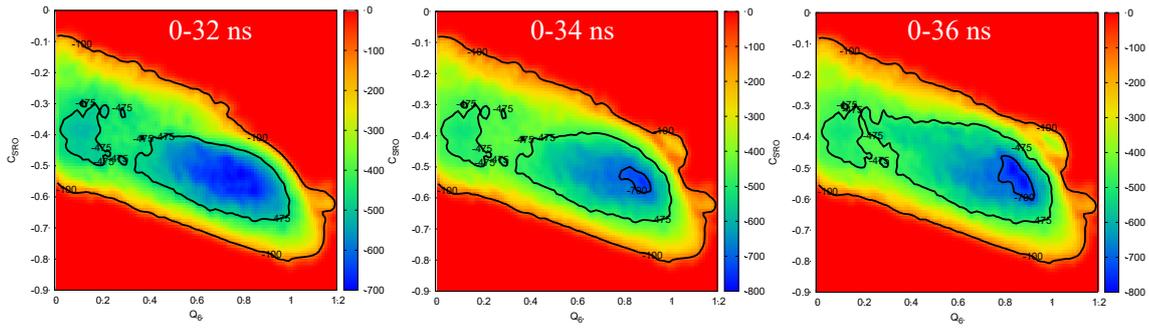

Fig. S2 The accumulated trajectories travelled on FES after every 2ns in [0, 36] ns region.

**VI. Structural evolution of the coordinate number and pair correlation function.**

We use coordinate number (CN) and pair-distribution function (PCF) to monitor the local structure change during the simulations. Figure S3 shows the CN distribution at I, II, IV, and V stages. Low CN is mainly dominated by Cu atoms, while the high CN is mainly dominated by Zr atoms. At the initial state, the CN of Cu and Zr atoms are no more than 13 and no less than 14, respectively. The local environments around Cu and Zr atoms are much different from the crystalline, because the local environment in the B2 crystal has a fourteen CN that eight atoms of the different type, compared with the central atom, are at the fist nearest neighbor (1NN) and six atoms of the same type are at 2NN. To achieve crystallization, the CN of Cu and Zr atoms should increase and decrease, respectively. Indeed, we observe the tendency that the number of CN-14 polyhedra is more than the number of non-CN-14 polyhedra at 9 ns, and the number of CN-14 polyhedra reaches the most at 22.66 ns.

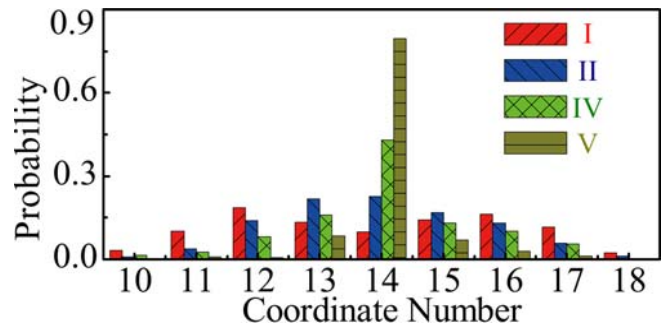

Fig. S3 Evolution for the distribution of polyhedra with different CN.

To further uncover the pathway of atom movement, partial PCFs of the four instantaneous stages are analyzed and shown in Fig. S4. We surprisingly found that the Cu−Cu PCF change the most among the three partial PCFs from the I state to the V state, whose first peak move from 2.5 Å to 3.5 Å. We found that the first peaks for Cu−Zr and Zr−Zr pairs change a little within 0.5 Å. In particular, it clearly shows that, at 9ns, the first peak of Cu−Cu partial PCF collapses while the first valley plumps up. We can predict that the movement of Cu atoms is the crystallization-determining step. The abnormal movement of $R_{pp}$ towards a large value before critical nucleation discussed in the main text, as shown in Fig. 1(a), originates from the special Cu movement to adjust the local chemical order, demonstrated in Fig. S3-S4. Finally, we can make a conclusion that the adjustment of the local environments, particular the Cu atoms around the Cu central atoms, is the key to achieve the rejuvenated disorder state and accomplish critical nucleation further.

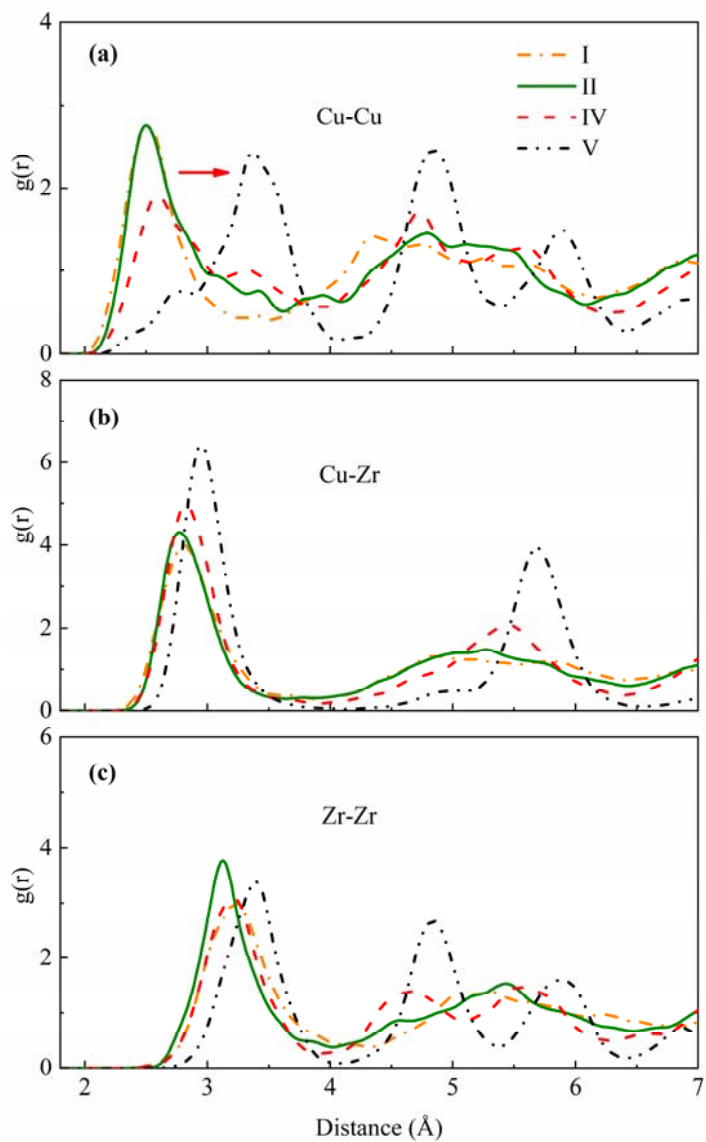

Fig. S4 Evolution of partial PCFs for Cu−Cu, Cu−Zr and Zr−Zr pairs, respectively. All the structures are obtained from the MTD trajectories of the I, II, IV, and V stages without relaxation.

**VII. The main polyhedral distributions in I and II states.**

Figure S5 shows the fraction of the main Cu-central polyhedral and Zr-central polyhedral by comparing the local structures of I and the II states after annealing 10ns at 300K. In our previous work [3], based on the analysis of strain energy, the most stability of Cu-center is (0,0,12,0) > (0,2,8,2) > (0,2,8,1) > (0,3,6,3) and the most stability of Zr-center polyhedra is (0,1,10,5) > (0,2,8,6) > (0,1,10,4) > (0,2,8,5). In the I state, the number of former three Cu-central polyhedral and the former three Zr-central polyhedral is larger than that in II-glass, and it is opposite for (0,3,8,3) and (0,2,8,5) polyhedral. It manifests that I-glass is more stable than II-glass. Therefore, we can make a conclusion that I-liquid has more third index of Voronoi index, representing more local fivefold symmetry (LFFS) [8] and more icosahedra-like.

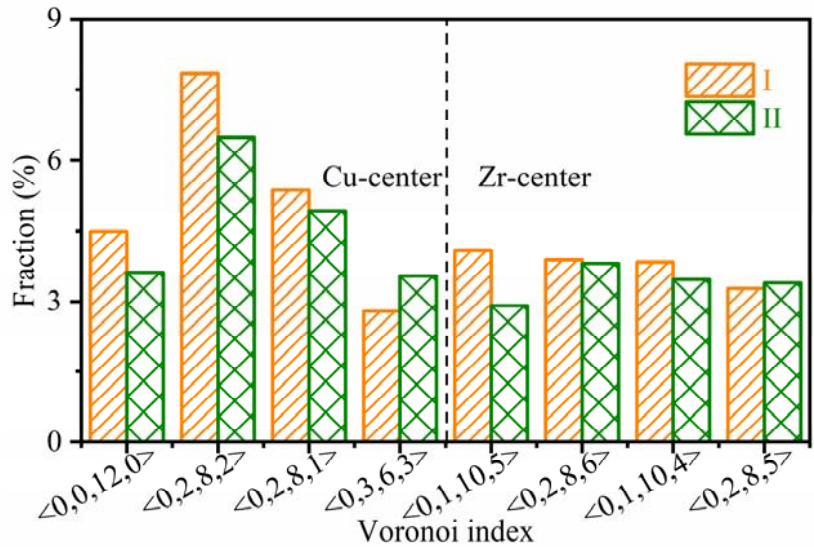

Fig. S5 The fraction of the main Cu-central polyhedral and Zr-central polyhedral by comparing the local structures of the I and the II states after annealing 10ns at 300K.

## VIII. The simulated FES of $Zr_{50}Cu_{50}$ at 600 K.

In order to investigate the crystallization behavior at 600K, we carry out many MTD simulations with various Gaussians biased potentials, but we do not detect a crystallization pathway, whose one FES is shown in Fig. S6. The encountered states locate at the amorphous region. This result without crystallization originates from the unable adjustment of local chemical order below the glass transition temperature ($T_g \approx 675$ K) that the diffusion barrier becomes much higher as the metallic system becomes denser at low temperature.

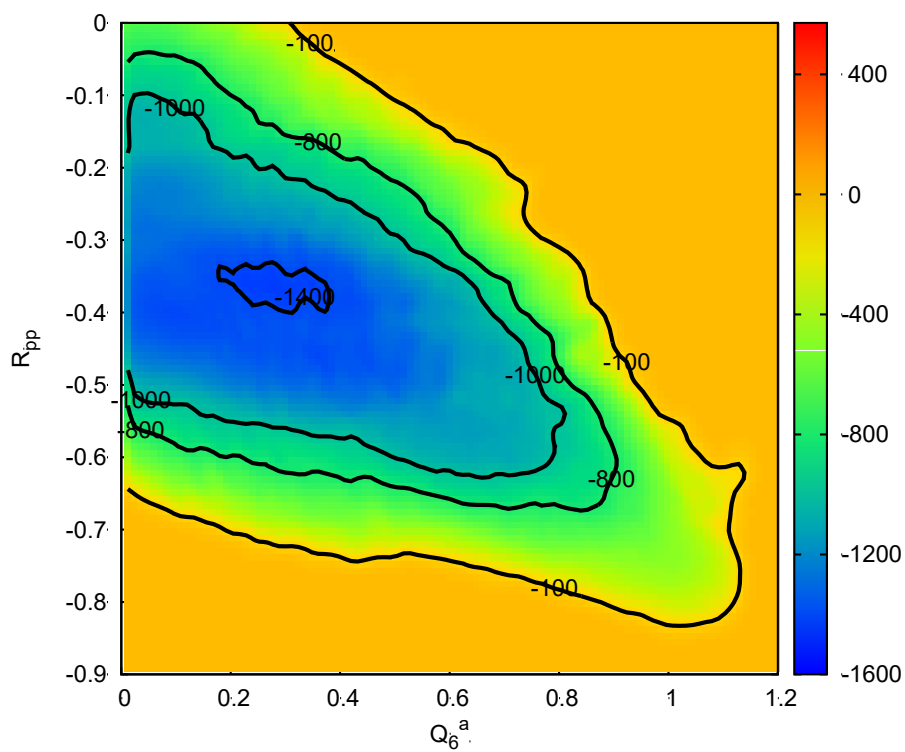

Fig. S6 The FES of $Zr_{50}Cu_{50}$ constructed using 75000 MTD steps (150 ns) at 600 K.

**IX. System size effect of metadynamics simulation.**

In order to investigate the size effect, we carry out a 2000-atoms $Zr_{50}Cu_{50}$ system at 800 K. The evolution of $R_{pp}$ and $Q_6$ CVs during the crystallization process is shown in Fig. S7. To our surprised, we find that the system crystallizes through many steps, instead of complete crystallization at one time. Particularly, two abnormal $R_{pp}$ evolution marked by two ellipses represents two obviously change of the local chemical order like the phenomenon in the 686-atoms system as shown in Fig. 1. We relax the two structures at I and II moments, where the former just has a small crystallite of 573 atoms and the latter is almost full crystalline atoms, as shown in Fig. S7(b)-(c).

The abnormal multistep crystallization originates from the much slow growth speed at the temperature near $T_g$, which is demonstrated in the reference [9]. Without the change of the local chemical order, the formed small crystallite cannot grow further, although the crystallite is stable to resist the melt. It illustrates that the nucleation and growth processes should encounter the rejuvenated disorder states. We also can make a conclusion that the 686-atom system choose in this work is large enough for the crystallization study in the deep undercooled temperature.

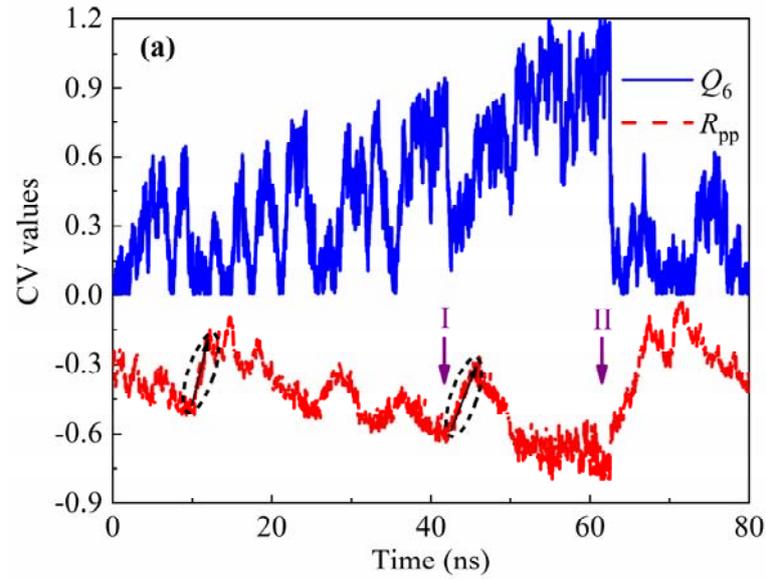

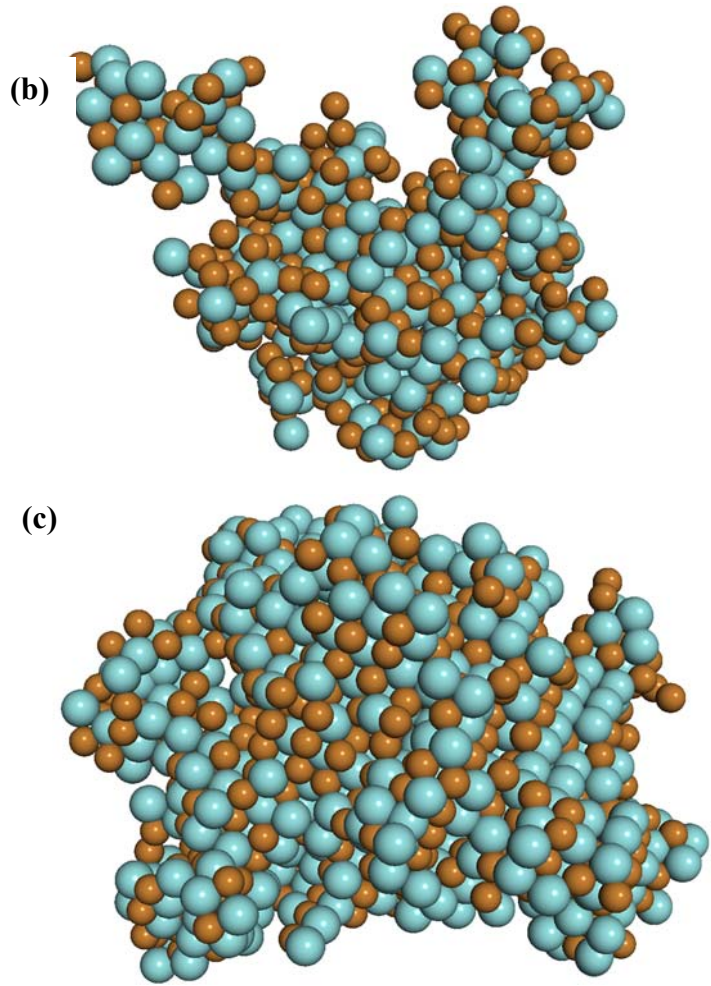

Fig. S7 The crystallization process of 2000-atoms $Zr_{50}Cu_{50}$ liquid at 800K. **(a)**, The evolution of two CVs, $R_{pp}$ and $Q_6$. **(b)-(c)**, The two snapshots of crystallites at II and II moments, respectively. The crystal-like atoms are selected by Voronoi index <0,6,0,8>.